\documentclass[reprint,amsmath,aps,prb,footinbib]{revtex4-2}
\UseRawInputEncoding
\usepackage{graphicx,epstopdf,dcolumn,bm,mathtools,siunitx,multirow,hyperref,xcolor,amssymb}
\newcommand{\mli}[1]{\mathit{#1}}
\begin{document}
\title{Origin of quantum shape effect}
\author{Alhun Aydin$^{1,2}$}
\email{alhunaydin@fas.harvard.edu}
\author{Altug Sisman$^{3}$}
\affiliation{$^{1}$Department of Physics, Harvard University, Cambridge, MA 02138, USA \\
$^{2}$Department of Physics, Ko\c{c} University, 34450 Sar\i yer, Istanbul, Turkey \\
$^{3}$Department of Physics and Astronomy, Uppsala University, 75120, Uppsala, Sweden}
\date{\today}
\begin{abstract}
Size-invariant shape transformation gives rise to the so-called quantum shape effect in strongly confined systems. While quantum size and shape effects are often thought to be difficult to distinguish because of their coexistence, it is actually possible to separate them and focus solely on the shape effect. In fact, quantum shape effect is a quite different phenomenon from quantum size effects, as it can have the opposite influence on the physical properties of nanoscale systems. Here we explore the origin of the quantum shape effect by theoretically investigating the simplest system that can produce the same physics: quantum particles in a box separated by a moving partition. The partition moves quasistatically from one end of the box to the other, allowing the system to remain in equilibrium with a reservoir throughout the process. The partition and the boundaries are impenetrable by particles, forming two effectively interconnected regions. Position of the partition becomes the shape variable. We investigate quantum shape effect on the thermodynamic properties of confined particles. In addition, we applied a new analytical model based on dimensional transitions to accurately predict thermodynamic properties under the quantum shape effect. A fundamental understanding of quantum shape effects could pave the way for employing them to engineer physical properties and design better materials at nanoscale.
\end{abstract}
\maketitle
\section{Introduction}
Quantum size effect (QSE) phenomenon is one of the pillars of nanoscience and nanotechnology that has been shaping modern technology for at least the last two decades. Reducing the sizes of a material to nanoscale causes them to exhibit size-dependent quantum-mechanical phenomena that are unseen at the macro scale \cite{rodun}. Existence of QSE is initially predicted by Herbert Fr\"{o}hlich in 1937 \cite{Fr_hlich_1937} and their effects on thermal properties are examined by Ryogo Kubo in 1962 \cite{Kubo_1962}. In nanoscale materials infinite-size approximations such as taking the thermodynamic limit fail \cite{baltes,molina,pathria}. Boundary effects and the geometry in which the particles are confined become significant when the sizes are reduced to nanoscale \cite{sando,bermanq,PhysRevLett.81.1062,dai1,sismanmuller,Dai_2004,sisman,doi:10.1063/1.2821248,flopb1,efspectra,Fp_2009,phystoday}. Quantum confinement makes discrete energy spectra to prominently reveal itself and, as a result, physical properties such as electronic \cite{Ciraci_1986,Erko__1987,Jal_ochowski_1988,bineker,doi:10.1126/science.262.5131.218,dattabook1995}, phononic \cite{Arora_1981},  magnetic \cite{Halperin_1986,Liu_1990,Azbel__2000,Yeh_2006,Zanjani2021,Koksharov}, mechanical \cite{Wu_2020}, optical \cite{Dryzek_1987,Yoffe_2002,PhysRevA.89.033835}, thermal \cite{Ozturk_2009,McNamara_2010}, thermodynamic \cite{Denton_1973,Volokitin_1996,Horodecki_2013,PhysRevLett.120.170601,modelnano,PhysRevE.101.012116,PhysRevE.97.042104}, thermoelectric \cite{tebook2013,tebook2014,Mao_2016} and superconducting \cite{Vopat_1987,suprqsesc,nanscsh} properties of materials become size dependent at nanoscale \cite{qconren,mitin,Victo-2014,GEOFFRION2020109320}.

Considering the importance and impact of size-dependent phenomena in today's science and technology, it is intriguing to ask about shape dependence and how does shape affect the physical properties of materials at nanoscale. While the sizes of a geometric object are defined by Lebesgue measure \cite{lebesgue}, shape is not so easy to readily characterize. When the phrase "shape effect" is used in physics literature, usually variations in boundary curvatures, anisometry (i.e. different aspect ratios), simple geometric structure (e.g. cubic, spherical) etc. are considered as indicators of the shape characteristics of a domain \cite{biss,shapemat,Glotzer2012,pchenprb,nantech,nanscsh,langmu,pccp1,PhysRevB.50.17721,shapeeffnl,PhysRevLett.118.157402,Potempa_1998,Scher_2016,khordad_effects_2018,PhysRevB.99.195303,PhysRevB.99.195447}. However, in those types of so-called "shape effects", sizes (either bulk or low-dimensional) of the objects also change along with shape because the care has not been taken to keep the geometric size variables constant during a shape transformation. In such a case, size and shape effects are inherently linked to each other. It is not possible to examine the effect of pure shape dependence without keeping the size parameters constant and preventing them to interfere with size effects.

A specific type of geometric transformation so-called the size-invariant shape transformation provides a way to keep the sizes of a domain fixed while continuously being able to change its shape \cite{aydin7}. By this way, size and shape effects can completely be separated from each other, allowing us to focus on how shape alone changes the physical properties of confined systems. For particles that are strongly confined (i.e. discrete energy spectrum becomes prominent) within a domain, any physical effect resulting from a size-invariant shape transformation is called quantum shape effect \cite{aydin7,aydinphd}. Note that quantum shape effect (QShE) requires strong confinement (in other words strong QSE) to appear in the first place. However, this does not mean that they have the same origin, as we will explore in this paper.

QShE has already been applied to nanoscale materials and conceptual devices. It has been shown that QShE can be used to design new devices for energy harvesting with nanostructures \cite{aydin11}. QShE makes it possible to construct quantum heat machines that are classically impossible \cite{aydin7,aydinphd}. Quantum oscillations induced by QShE have been predicted in the thermodynamic properties of core–shell nanostructures, providing a new mechanism for changing the polarity of semiconductors and fine tuning their Fermi level \cite{aydin12}. QShE phenomenon has also implications in quantum Szilard engines, where it modifies the values of heat and work exchanges \cite{aydin10}. 

The aim of this article is to explicitly and comprehensively explain the origin and the mechanisms of the QShE phenomenon. To this end, we present the simplest possible system that is able to exhibit QShE: particle in a box with moving partition. In particular, we consider non-interacting particles obeying Maxwell-Boltzmann statistics confined in a box with a movable partition. Note that we deliberately simplified our system as much as we can by preserving the roots of the effect. Besides the numerical calculations, we also applied recently proposed analytical methods \cite{aydin2,aydin7} both to accurately calculate the partition function of the system, and to clarify the origin of the quantum shape effect. Finally, we investigate the unconventional behaviors of thermodynamic properties under quantum shape effect. 

\section{The origin of quantum size effect}

In order to describe the origin of the QShE, first we need to discuss the origin of the QSE. In this section, we describe the physical mechanism leading to QSE. Then we explicitly define the sizes of a material. We discuss several methods to get analytical expressions for QSE corrections in statistical mechanics by taking 1D partition function as the simplest example. Finally, we mention the strong confinement case, where the discrete energy levels play an explicit role on determining QSE. After building up a clear understanding for QSE, it will be easier to describe the QShE and their difference from QSE. Note that we do not consider classical size effects, classical shape effects or system size effects (i.e. number of the constituents of the system) in this article. Our attention will be given to quantum size and quantum shape effects.

\subsection{Size quantization: Reducing sizes to a few thermal wavelengths}
Discreteness of (or spacing between) the quantized energy levels of particles for quadratic dispersion relation is given by the relation $\Delta E \sim\hbar^2/(mL^2)$ where $\hbar$ is Planck constant, $m$ is the mass of a particle and $L$ is the size of the material. Similarly, the typical thermal energy scale of particles is $E_{th} \sim k_BT\sim\hbar^2/(m \lambda_{th}^2)$ where $\lambda_{th}=\hbar\sqrt{2\pi}/\sqrt{mk_BT}$ is the thermal de Broglie wavelength of particles (corresponds to the average size of the space that particles occupy), which is in the order of nanoscale for practical temperatures. At macroscale, discreteness of energy levels become negligible compared to the thermal energy of particles, i.e. $\Delta E<<E_{th}$. Therefore, energy levels of particles in bulk materials at macroscale can safely be considered as continuum. In other words, one can fit many thermal wavelengths into a large domain $(L/\lambda_{th}>>1)$, so that huge number of modes can be thermally excited. 

Reducing the sizes of the materials to the order of the thermal de Broglie wavelengths of particles, $L\approx\lambda_{th}$, causes energy level spacings to increase ($\Delta E \sim 1/L^2$) which makes the discrete energy levels to reveal themselves, $\Delta E\approx E_{th}$. Thermal energy can only excite a few lower energy levels. From another perspective, only a few number of thermal wavelengths can fit into such a small space, $(L/\lambda_{th}\sim 1)$, see the comparison in Fig. 1(a) and (b). Due to this reason, QSE is also called size quantization. When the size of a domain is that small, particles can no longer be treated as point-particles as their wave nature becomes prominent. Discreteness of the spectrum plays an important role since only a few low-lying energy levels can be thermally occupied. Comparison of the thermal occupation probabilities of quantum states can be seen for two domains with different lengths in Fig. 1(c). Revelation of the degree of discreteness (or quantization) of energy levels due to prominent wave nature of particles makes the physical quantities dependent on additional functions of the system sizes, which is \textit{the} origin of the quantum size effect. As a matter of course, both wave nature and discrete spectrum are the properties of particles even at macroscale. But QSE is actually related to the fact that these properties becoming prominent at nanoscale due to the increased influence of the ground and low-lying states which dominates the behaviors of the physical properties and causing nanoscale systems to exhibit considerably different behaviors than the bulk materials.

\begin{figure}[t]
\centering
\includegraphics[width=0.48\textwidth]{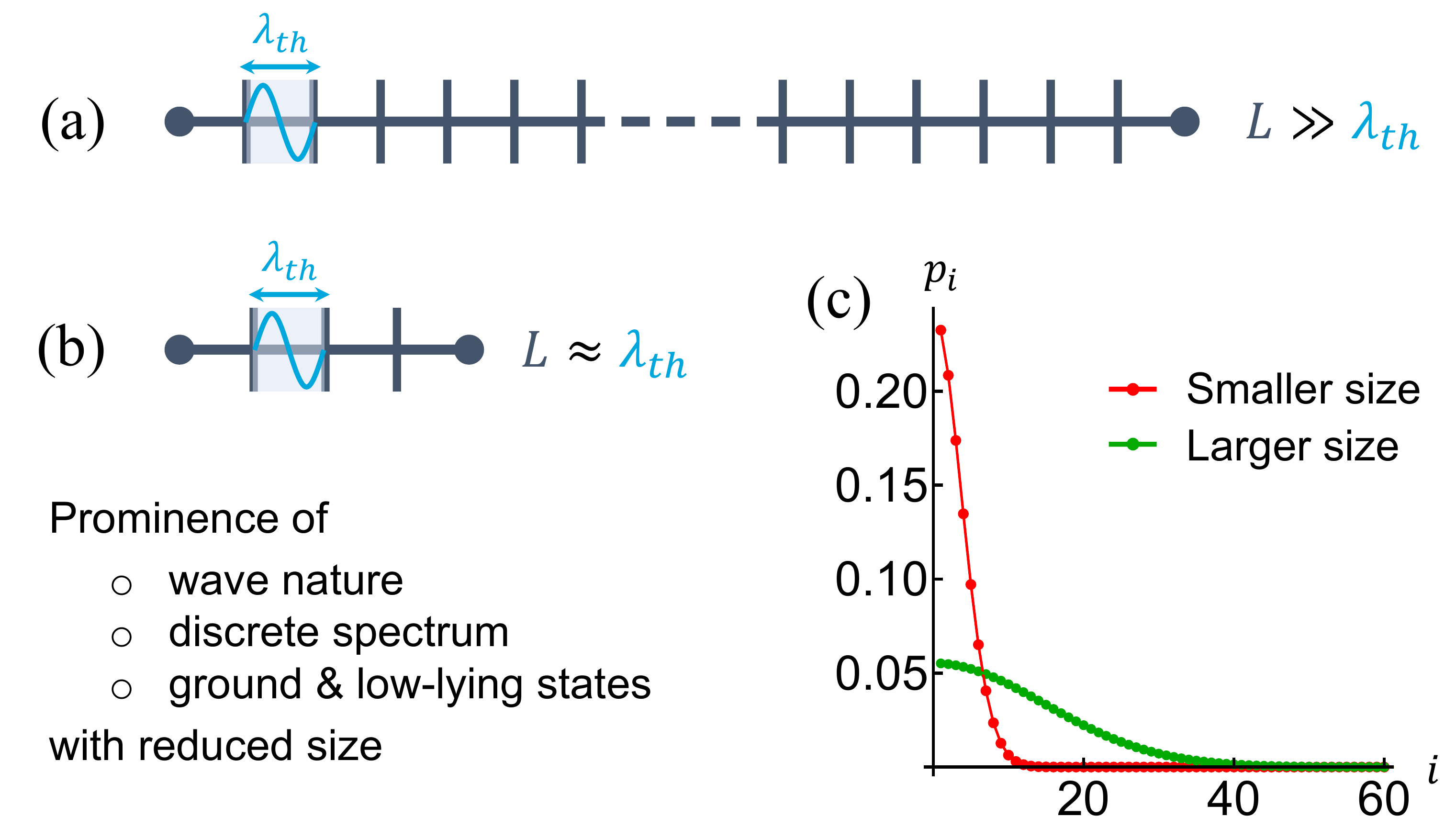}
\caption{The origin of quantum size effect. (a) Domain length is much larger than the thermal de Broglie wavelength of particles so that many wavelengths can fit into the domain. (b) Domain length consists of a few thermal wavelengths so that the wave nature, ground state energy and discrete energy spectrum of particles become appreciable. (c) Thermal probability (Boltzmann) of states for two different cases. For the smaller domain size, discrete spectrum becomes prominent and only a few low-lying states are excited. For larger domain size, discrete spectrum starts to become negligible and higher excited states contribute.}
\label{fig:pic1}
\end{figure}

While the discrete spectrum becomes prominent at nanoscale, considering a continuous spectrum bounded from the below will be practically enough to generate all the QSE corrections, except for the case of strong QSE \cite{aydin3}. As we mentioned, when the sizes of the system becomes close to the thermal de Broglie wavelength of particles, the system could be accurately described by fewer and fewer modes that are thermally populated. Due to this "soft cutoff" provided by the relevant thermal distribution, ground state contribution becomes appreciable. This makes it possible to obtain QSE corrections by neglecting the discreteness of the spectrum and just by correctly accommodating the ground state contribution, as we shall see in the following subsections.

Quantum confinement giving rise to QSE is basically an inverse measure of how many thermal wavelengths can fit into a given domain length. Due to this, it is sometimes also called size quantization (size itself is not quantized like the energy is). In this regard, QSE can be quantified by defining confinement parameters \cite{aydin1}, which are basically the ratios of two characteristic length scales of the system, e.g. $\alpha\sim\lambda_{th}/L$. Then many physical properties, such as mechanical, electrical, optical and thermal properties, of materials become explicit functions of confinement parameters. Additionally, strong confinement in one or more directions introduce the aspect of dimensionality (e.g. low-dimensional structures) and materials start to exhibit drastically different behaviors as a result of radical variations in the behavior of density of states \cite{rodun,mitin}.

\subsection{Geometric size variables}
In the previous subsection, we used a generic parameter $L$ as the size of a system just to be able to make comparison between the sizes of macro and nano worlds. In fact, size has a well-defined meaning and description in mathematics. In measure theory, the sizes of a geometric object are described by Lebesgue measure \cite{lebesgue}. Sizes of a 3D object are determined by its volume $\mathcal{V}$, surface area $\mathcal{S}$, periphery $\mathcal{P}$ and number of vertices (i.e. any kind of boundary discontinuities such as cusps, corners or dots in 1D) $\mathcal{N_V}$. These parameters are called geometric size parameters. For lower dimensional objects, the same procedure applies with the reduction of dimensions, see Fig. 2. Volume, surface area and periphery of an object might be more obvious, whereas the number of vertices can be tricky to calculate especially for arbitrary domains \cite{aydinphd}. Let's mention for clarity that the number of vertices of the objects in the column III of Fig. 2 are 4, 3 and 3 from top to bottom respectively.

\begin{figure}[t]
\centering
\includegraphics[width=0.35\textwidth]{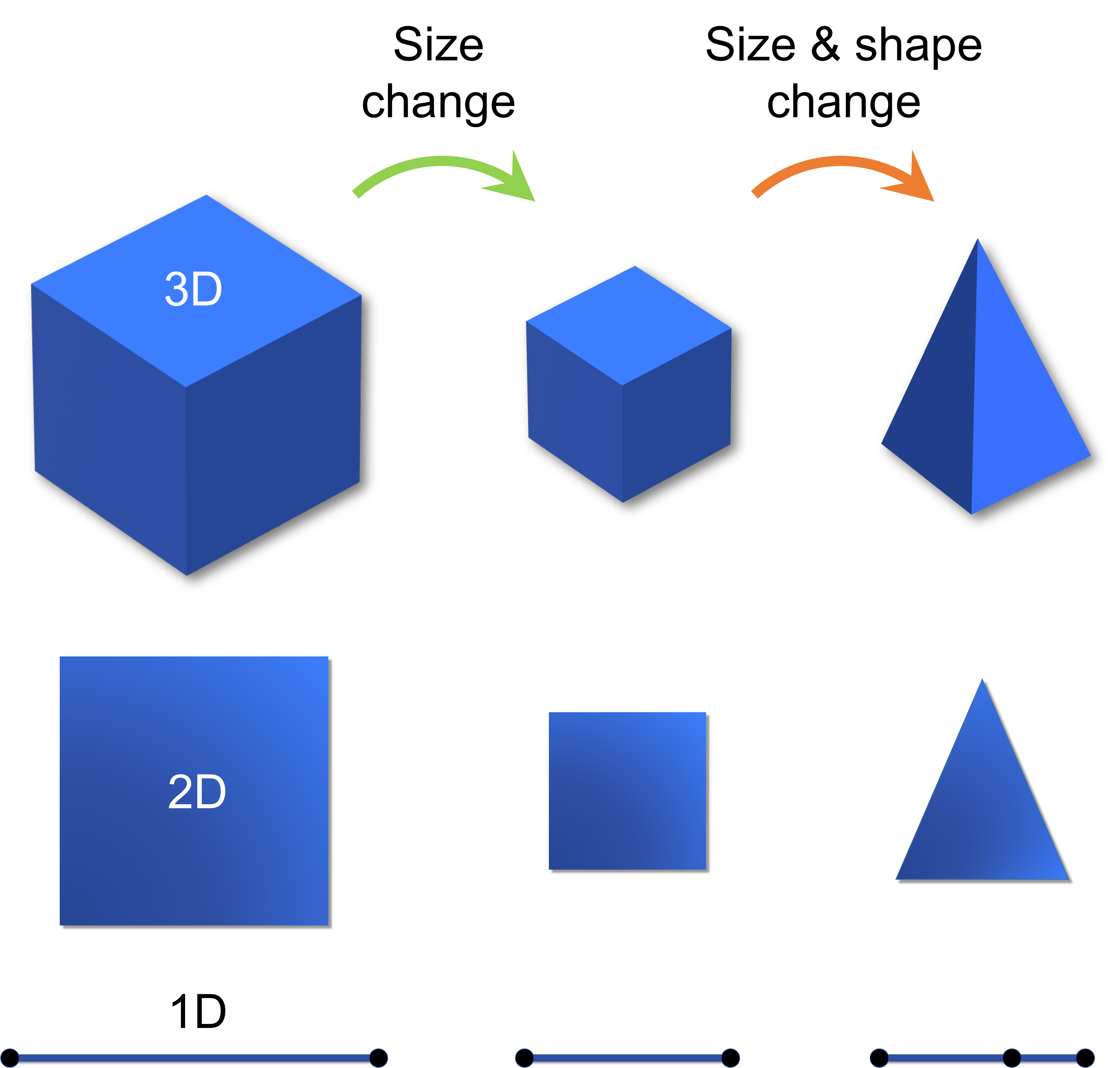}
\caption{Changing sizes in various dimensions. (Top row) Larger cube, smaller cube and tetrahedron, (middle row) larger square, smaller square and triangle, (bottom row) longer line, shorter line and a line with an additional inner boundary. Sizes are defined under the standard Lebesgue measure. A 3D domain is characterized by four different geometric size variables: volume $\mathcal{V}$, surface area $\mathcal{A}$, peripheral lengths $\mathcal{P}$ and the number of vertices $\mathcal{N_V}$. Similarly, 2D domain is by $\mathcal{A, P, N_V}$ and 1D by $\mathcal{P, N_V}$. Sizes of an object can be changed while keeping its general shape constant by uniform scaling (comparison of first and second columns). When second and third columns are compared, it can be seen that not only the size but also shape of the objects have changed.}
\label{fig:pic2}
\end{figure}

Geometric size parameters fully describe the sizes of a material. Most of the time the leading order QSE correction is sufficient, however, lower order corrections can also become appreciable depending on the size and geometry of the domain that is considered. To change the size of a material, one needs to change at least one of the geometric size parameters. On the other hand, if all geometric size parameters are the same for two materials, then their sizes are equal and QSE corrections on material properties are exactly the same for both materials.

\subsection{Analytical methods to obtain quantum size effects}

There are several convenient methods to get QSE correction terms to the usual statistical mechanical expressions. Here, we present them using a simple example. Consider non-relativistic quantum particles confined in a one-dimensional box with length $L$. The boundaries are assumed to be impenetrable, i.e., the potential is zero within the box and infinite on the boundaries. The domain length $L$ determines the quantized energy levels of particles confined inside the box. One can write down the partition function and then calculate the thermodynamic quantities for the particles confined in the box within the standard statistical mechanical framework.

For a canonical ensemble, the exact form of the partition function contains the summation of Boltzmann factors over the discrete energy spectrum $E_i$ with state variable $i$. If the domain length $L$ is large enough, one can use continuum approximation (CA) for energy levels and replace the summation operator with integral to get the textbook expressions of the single-particle partition function as
\begin{equation}
\begin{split}
\zeta=\sum_i\exp(-\beta E_i)\xrightarrow{\text{CA}}\,\,\,\, \approx\int\exp(-\beta E_i)di,
\end{split}
\end{equation}
where $\beta=1/(k_BT)$ with Boltzmann constant $k_B$ and temperature $T$. For a particle confined in a 1D domain, $f(i)=\exp\left[-\frac{\pi}{4}\left(\frac{\lambda_{th}}{L}i\right)^2\right]$, continuum approximation gives
\begin{equation}
\begin{split}
\zeta =\sum_{i=1}^{\infty}f(i)\approx\int_{0}^{\infty}f(i)di=\frac{L}{\lambda_{th}}
\end{split}
\end{equation}

In 1D, volume of the system is basically its length, $L$, and the partition function is just the ratio of domain size and thermal de Broglie wavelength of particles. In that sense, partition function tells us how many thermal wavelengths we can fit into our domain. On the contrary case, if $L$ is small enough so that the thermal de Broglie wavelength of particles is in the order of $L$, QSE (or finite-size effects) become important and conversion from summation to integration should be avoided. The summation over discrete energy levels fully takes any QSE into account, however their analytical relations with size are implicit within the sum. It is worthwhile to notice that the single-particle partition function itself serves as a measure of QSE and size-quantization. When $\zeta\approx 1$, QSE and quantization are strong, whereas if $\zeta>>1$, QSE and quantization are negligible. This fact is also reflected in the definition of confinement parameters, which up to some numerical factor is the inverse of the single-particle partition function.

In order to explicitly (i.e. analytically) examine QSE, one needs to make a better approximations than the continuum one. There are three effective ways of obtaining the analytical QSE corrections to the continuum expressions: (1) Poisson summation formula (2) Weyl density of states and (3) quantum boundary layer method. While all of them give the same correct answers, each method contributes to the mathematical and physical understanding of QSE in their own way. In order to distinguish quantum shape effects from the size effects, it is essential to mention these methods which help to understand where the QSE come from.

\subsubsection{Poisson summation formula}

Replacing the summation directly with the integration is a coarse approach. One can represent the summation more precisely by applying the Poisson summation formula (PSF) \cite{baltes,path1,path2} or other summation formulas like Euler–Maclaurin formula \cite{Berry_2020,stattherm21}, and Abel-Plana formula \cite{Saharian_2009}. The full form of PSF for even functions reads
 \begin{equation}
\sum_{i=1}^{\infty}f(i)=\int_0^\infty f(i)di-\frac{f(0)}{2}+2\sum_{s=1}^{\infty}\int_0^\infty f(i)\cos(2\pi si)di.
\end{equation}
The first two terms of PSF are enough to generate the QSE terms which gives the following result,
\begin{equation}
\begin{split}
\zeta =\sum_{i=1}^{\infty}f(i)\approx\int_{0}^{\infty}f(i)di-\frac{f(0)}{2}=\frac{L}{\lambda_{th}}\left(1-\frac{\lambda_{th}}{2L}\right)
\end{split}
\end{equation}
The first term (integration) is the bulk term which is basically the continuum expression. The second term ($-f(0)/2$), on the other hand, generates the QSE correction term (the second term within the parenthesis in Eq. (4)). Approximating the sum with the first two terms of PSF is also called the bounded continuum approximation, because it takes into account the finiteness of the boundaries while still approximating energy levels as continuum. As it happens, this operation is not only a convenient mathematical approach for a summation, but also has a well-defined physical meaning as well. In other words, the second term in Eq. (3) corrects the false contribution of zeroth quantum state in the integral, which does not exist in a quantum mechanical energy spectrum. Remarkably, QSE corrections are directly determined by the correct treatment of the ground state contributions, which is basically removing the false contribution of zero ground state energy in the continuum approximation. This happens because the ground state mode of a confined direction represents the $(D-1)$ dimensional contribution in a $D-$dimensional domain. For example, here in 1D system, the contribution of the boundary in a 1D domain represents 0D, i.e. number of vertices, contributions. QSE, therefore, can be understood as the lower-dimensional boundary corrections to the higher dimensional bulk terms. This will become much more clear in next subsection. The third term of PSF, on the other hand, is called the discrete correction term as it fully recovers the discrete summation by correcting the miscalculations for each state in the continuum representations of the actual summation. 

Note that when $L>>\lambda_{th}$, QSE correction becomes negligible compared to unity (see Eq. 4) and the classical expression can be recovered. QSE corrections loose their meaning after $L \leq\lambda_{th}/2$, since one cannot use $\lambda_{th}$ beyond that limit as it is defined based on a full wavepacket in an unbounded domain. For references on the usage of PSF or related formulas in QSE, see \cite{sismanmuller,Dai_2004,aydin1}.

\subsubsection{Weyl density of states}

Another method to obtain QSE corrections is using the Weyl density of states (WDOS) \cite{baltes,pathbook,qforce,aydin3,aydin8}. WDOS is a concept based on the Weyl law (also called Weyl conjecture) which describes the asymptotic behavior of the eigenvalues of a Laplacian \cite{weyl11,weylcomput20}.

Based on Weyl law, WDOS in energy space can be derived in its $D$-dimensional general form as \cite{aydinphd}
\begin{equation}
\begin{split}
W_D(E)=& \frac{\mathcal{V}}{\lambda_{th}^3}\frac{2\sqrt{\beta E}}{\sqrt{\pi}}\Theta(D-3)+(-1)^D\frac{\mathcal{A}}{\lambda_{th}^2}\frac{1}{4^{D-2}}\Theta(D-2) \\
& +(-1)^{D-1}\frac{\mathcal{P}}{\lambda_{th}}\frac{1}{4^{D-1}}\frac{1}{\sqrt{\pi}\sqrt{\beta E}}\Theta(D-1) \\
& +(-1)^{D-2}\frac{\mathcal{N_V}}{4^D}\delta_{\mli{Drc}}(\beta E)
\end{split}
\end{equation}
where $D$ is the dimensionality, $\Theta$ is right-continuous Heaviside step function, $\delta_{\mli{Drc}}$ is Dirac delta function, decorated with subscript $\mli{Drc}$ in order to prevent confusion with the parameter that will be introduced in the next subsection.

Instead of using the usual density of states expressions one can use WDOS inside the integration which generates lower dimensional geometric corrections to the expressions in addition to the bulk term. An example of the usage of WDOS to obtain the QSE in partition function is given below,

\begin{equation}
\begin{split}
\zeta =\sum_{i=1}^{\infty}f(i) \approx\int_{0}^{\infty}f(E)W_1(E)dE
=\frac{\mathcal{P}}{\lambda_{th}}\left(1-\frac{\lambda_{th}}{4}\frac{\mathcal{N_V}}{\mathcal{P}}\right)
\end{split}
\end{equation}
where $\mathcal{P}$ is periphery of the domain, which is actually the length of the domain in 1D, so $\mathcal{P}=L$ and $\mathcal{N_V}$ is the number of vertices, which is 2 (left and right boundaries) and WDOS in 1D is $W_1(E)=(L/\lambda_{th})/(1/\sqrt{\pi\beta E})-(\mathcal{N_V}/4)\delta_{\mli{Drc}}(\beta E)$. As is seen, Eqs. (4) and (6) are equivalent to each other. At nanoscale systems, WDOS gives much more accurate results than the conventional DOS. Note that the domains that are designed to give QShE would have exactly the same asymptotic Weyl spectrum, since all Weyl parameters are exactly the same up to the lowest order. Therefore, WDOS cannot predict QShE. For references on the explicit usage of WDOS, see \cite{aydin3,aydin8}.

\subsubsection{Quantum boundary layer method}

Quantum boundary layer (QBL) method provides another (physically more intuitive) way that can effectively be used to analytically derive the correct forms of QSE corrections to the continuum expressions \cite{qbl,uqbl,nanocav}. In addition, QBL method reduces a quantum mechanical problem, QSE, to a geometrical problem.

The essence of QBL method relies on the quantum-mechanical particle density distribution profile which carries crucial information regarding the confinement. At thermal equilibrium, local density distribution of particles confined in a domain is described by ensemble-averaged quantum-mechanical particle number density, or in short quantum thermal density, which is given by
\begin{equation}
\begin{split}
n(\mathbf{r})=\sum_{\mathbf{k}}f(E_\mathbf{k})\left|\psi_\mathbf{k}(\mathbf{r})\right|^2
\end{split}
\end{equation}
where $\mathbf{k}$ represents the generalized quantum state variable and $\psi_\mathbf{k}(\mathbf{r})$ is the eigenfunction corresponding to the eigenvalue $\mathbf{k}$. For a particle in a 1D box, $\psi_i(x)=\sqrt{2/L}\sin(i\pi x/L)$ where $i$ is the quantum state variable. Due to the wave nature of particles, the density distribution profile of the particles is non-uniform even at thermal equilibrium, see the black curves in Fig. 3. This non-uniformity is a direct consequence of the fact that a few low-lying eigenstates are thermally occupied so that low-lying eigenfunctions dominate the local density behaviors. For macroscopic systems, non-uniform density distribution is negligible, whereas at nanoscale it becomes appreciable. It turns out that the degree of non-uniformity is directly related with the magnitude of QSE, which we can exploit to analytically obtain QSE corrections. 

As is seen from the black curves in Fig. 3, the density distribution of particles has two distinct regions, the central plateau (flat density region) and boundary layers (the decay of density near the boundaries). Also see the second row in Fig. 3 for the same physics in 2D domain. QBL method approximates this density distribution by considering a uniform maximal density region in the center and completely empty regions near the boundaries, see dashed red curves in Fig. 3. In other words, within the QBL approach, particles are assumed to only occupy an effective region that is described by the uniform density part, rather than occupying the whole domain $L$. Uniform density region is called the effective size (e.g. effective volume in 3D, or effective length, $L_{\mli{eff}}$ here in 1D) and empty regions are called quantum boundary layers. The height of the plateau region (maximum density value) determines the thickness of the QBL, since the domain integral of dimensionless density ($n/n_{cl}$ where $n_{cl}=N/\mathcal{V}$, the number of particles divided by the apparent volume) has to be equal to unity due to probability conservation. For Maxwell-Boltzmann distribution function, the thickness of QBL has been found as $\delta=\lambda_{th}/4$, which is independent of geometry and dimensionality \cite{uqbl}. Then the effective length can be expressed as $L_{\mli{eff}}=L-2\delta$.

\begin{figure}[t]
\centering
\includegraphics[width=0.48\textwidth]{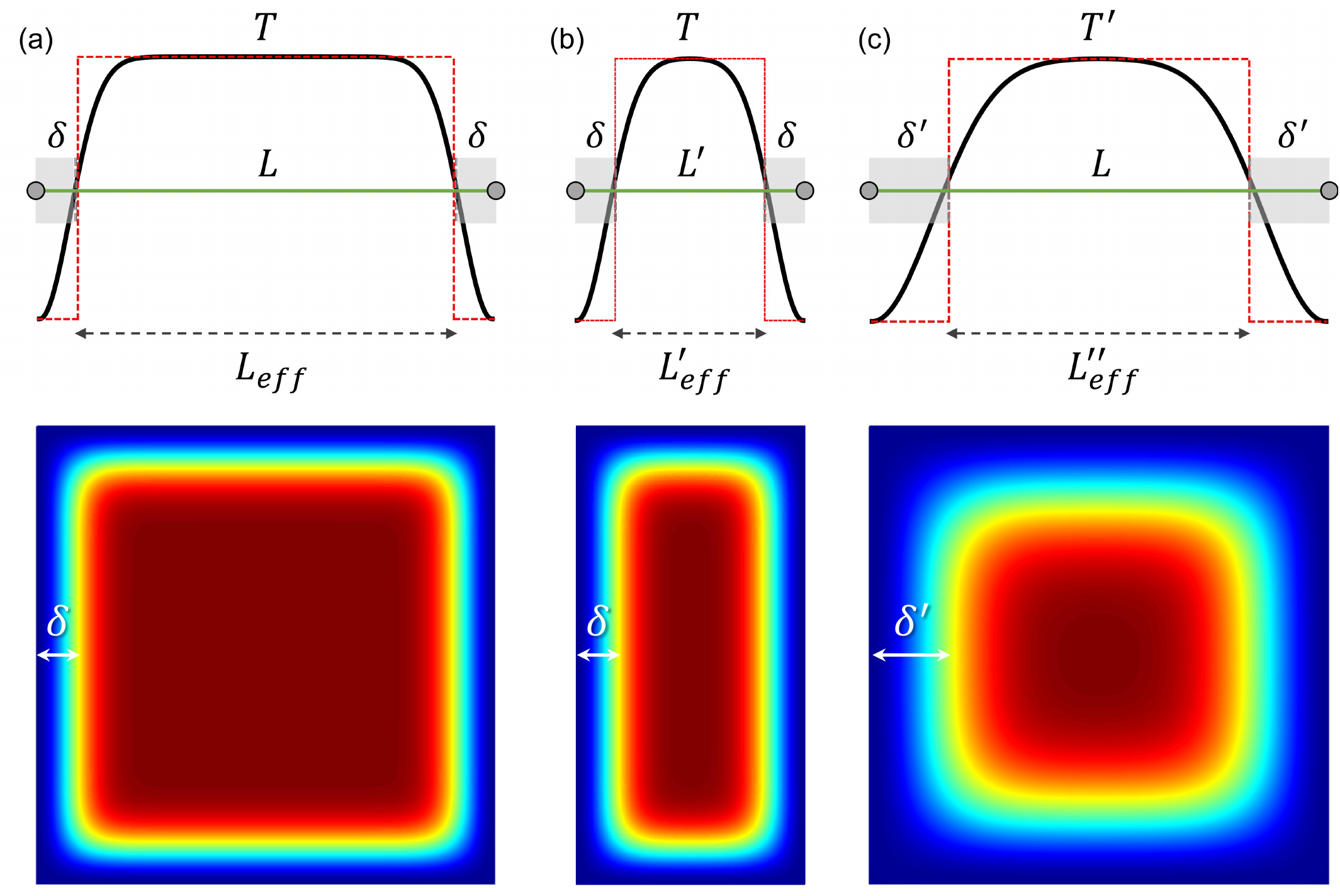}
\caption{Formation of quantum boundary layers and their variation with respect to changes in domain size or equilibrium temperature of particles. Green lines are actual domain lengths restricted by impenetrable domain boundaries denoted by grey dots. Dotted black arrowed lines show effective length perceived by the particles as a consequence of quantum size effects. Solid black curves represent the quantum thermal density distributions of particles inside a 1D domain in an arbitrary scale (for upper row, $x-$axis is the position coordinate and $y-$axis is the quantum thermal density). Dashed red curves indicate the approximation carried out by QBL method to represent the exact density distributions. Grey areas denote the thickness of QBL, $\delta$, which is a function of temperature. From (a) to (b) the domain size is reduced. $\delta$ remains the same, however quantum size effects increase because QBLs comprise larger portion of the actual domain size in (b) compared to (a). From (a) to (c) the domain size is kept constant whereas the temperature is decreased. $\delta$ increases with decreasing temperature which again leads to increase in quantum size effects due to the same reason with a different cause. Second row illustrate the same in a 2D domain where $x-$ and $y-$ axes are coordinates and the rainbow color scheme denotes the density so that the darkest blue and red represent zero and maximum densities respectively.}
\label{fig:pic3}
\end{figure}

Now, QSE corrections can be obtained just by replacing the length $L$ with the effective one, $L_{\mli{eff}}$, in the bulk expression. Then using QBL method, the same expression can immediately be recovered as 
\begin{subequations}
\begin{align}
\zeta =\sum_{i=1}^{\infty}f(i) & \approx \frac{L_{\mli{eff}}}{L}\int_0^\infty f(i)di \\
& \approx\frac{L_{\mli{eff}}}{\lambda_{th}}=\frac{L-2\delta}{\lambda_{th}}=\frac{L}{4\delta}\left(1-\frac{2\delta}{L}\right)
\end{align}
\end{subequations}
The QBL thickness is $\delta=\lambda_{th}/4$ for particles obeying Maxwell-Boltzmann statistics, so Eqs. (8) are also equivalent to Eqs. (4) and (6). We want to stress that this representation is not just a conventional trick happens to work by chance. Recall that partition function can be interpreted as the number of thermal wavelengths that can fit into a domain. However, in order to fit thermal wavelengths into a domain, those parts of the domain must be available for particles to occupy in the first place. In strong confinements, the wave nature of particles and the impenetrable nature of the boundaries cause a sharp decrease in the spatial occupation probabilities near the boundaries. This near-zero occupation space has an average thickness of $\delta$ and the effective length concept basically amounts to removing these parts from the actual length of the domain so that partition function can now be defined as the number of wavelengths that can fit into the effective (occupiable) length of the domain.

QBL method provides additional physical insights to the QSE phenomenon. For example, QSE increases when domain sizes are reduced, because QBLs form a larger portion of the whole domain, leaving effectively less space for particles to occupy, Fig. 3. This can be achieved either by changing the domain sizes (compare Fig. 3(a) and (b)) or by changing the temperature (compare Fig. 3(a) and (c)) or using particles with some other mass. QBL method also explains the reason why correction terms add up with alternating sign or why do they appear in that form \cite{qbl,aydinphd}. Furthermore, the necessary work to evacuate QBL region exactly equals to the QSE term appearing in the free energy expression. In this way, QBL provides physical explanations for each QSE term in thermodynamic properties as well as explaining their mathematical and geometrical origins.

Besides providing physical understanding, QBL method generates the QSE corrections without needing to explicitly solve the Schrodinger equation and practically leaving out the burden of calculating the summations. It has been also shown that QBL method gives accurate estimates even in arbitrary-shaped domains \cite{aydinphd}. For references on the usage of QBL method in QSE, see \cite{qbl,uqbl,nanocav}. For a more detailed analysis of the method, see Section 2.3 of Ref \cite{aydinphd}.

It should be noted that we used 1D results and Maxwell-Boltzmann statistics to simply demonstrate the methods. Further details and the generality of the methods are beyond the scope of this article. Nonetheless, all three methods are directly applicable to and works well in any dimension as well as in Fermi-Dirac and Bose-Einstein statistics.

\subsection{Quantum size effects beyond Weyl terms: Strong confinements}

All three methods that are presented in the previous subsections contain certain approximations within them. In the first method, the third term of PSF is neglected. Likewise in the second method, the error term in Weyl law is omitted. The third method relies on the zeroth order (stepwise) approximation of the QBL approach. Although in principle it should be possible to obtain arbitrarily accurate results by considering the higher order QBLs, this has not been accomplished yet due to the complexity of the geometric problem, especially in higher dimensions \cite{aydin7,aydinphd}.

All three methods manifest the bounded continuum approximation and still does not take the discreteness of energy levels into account. Therefore, they are only applicable to the weakly or moderately confined systems. Under stronger confinements, one cannot use these approximate methods and in most cases there is nothing to do but calculate the summations directly (where truncating after a few terms would be suitable). Nevertheless, in the recent years, new approaches have been developed and some efforts have been done on getting analytical results even under strong confinements \cite{aydin4,aydin5,aydin6,aydin7}. These strong confinement effects also give rise to interesting results such as discrete density of states \cite{aydin3}, dimensional transitions \cite{aydin2,PhysRevE.104.054110} and the intrinsic discrete nature of thermodynamic properties in Fermi gases \cite{aydin1}.

A tempting question to ask here could be whether these neglected terms can be attributed to QShE rather than QSE. Although it is true that QShE cannot be represented by the first two terms in the analytical approximations (because they are invariant under QShE), this does not mean that QShE is just due to the neglected terms, as we will show explicitly in Sec. III(C).

\section{The difference between quantum size and shape effects}

Quantum size and shape effects are generally considered to coexist and in most systems they are indeed. However, this does not mean that their influence on the system properties are similar, or they cannot be separated. In this section, we introduce pure QShE and show how they are fundamentally different from quantum size effects.

\subsection{Size-invariant shape transformation}

Shape is defined as the geometric information that is encoded by the boundaries, and invariant under translation, rotation and uniform scaling \cite{shape1,shape2}. QShE is a phenomenon that is caused solely by changing the shape of the system while keeping all other variables such as particle density, temperature, size, external fields (if applicable) constant. In the previous section, we have explicitly stated the sizes of an object by the Lebesgue measure. The question then is: Is it possible to change the shape of a domain by keeping all the geometric size parameters constant? It has been recently shown that this is possible by what is called a size-invariant shape transformation \cite{aydin7}, that is illustrated in Fig. 4.

\begin{figure*}[t]
\centering
\includegraphics[width=0.98\textwidth]{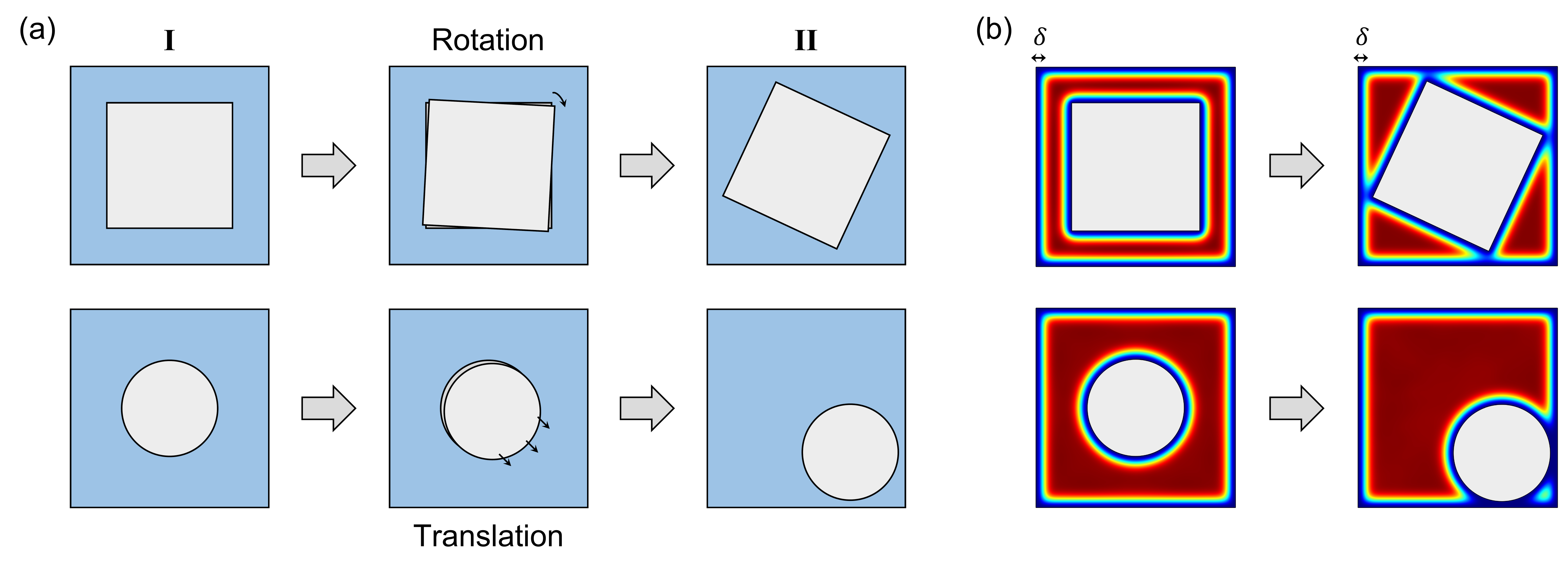}
\caption{Size-invariant shape transformation. (a) Consider two nested domains (e.g. square within a square in the top row or disk within a square, bottom row) where the outer (blue) part is the confinement domain, i.e. the domain where the particles are confined, whereas the inner (gray) part has infinite potential so that particles cannot penetrate into it. Applying rotation or translation to the inner object creates different confinement domains without changing any of the geometric size variables. Using this simple technique on nested domains, one can generate arbitrarily many domains with distinct shapes but having exactly the same sizes. (b) Quantum thermal density distributions of particles taking the shape of the domain that they are confined within. Quantum boundary layers (almost zero density regions) form near the impenetrable domain boundaries. (Rainbow color scheme is used: Red is higher, blue is lower density) Red regions designate the effective domain that particles can occupy at thermal equilibrium. We do not give a legend as the figures are for illustration purposes. Quantum boundary layers of inner and outer boundaries overlap when they come close to each other. This overlap carries important as well as analytically extractable information about the effective domain shape perceived by the particles.}
\label{fig:pic4}
\end{figure*}

Consider nested domains (a domain within a domain), like the ones in Fig. 4(a), where the outer domain is fixed and inner one is free to rotate. Outer domain (blue) is the one that particles are confined within, whereas the inner domain (gray) is impenetrable by the particles. Their initial configuration is denoted by (I). Now let's perform a rotation (top row) or translation (bottom row) to the inner domain. The resulting configurations become the ones given in (II). Once we compare the domains in (I) and (II), we can see that the shape of the blue region where the particles are confined changes under this transformation, while the sizes (all geometric size parameters) stay constant. Despite the Weyl terms are constant and the resulting spectrum has the same asymptote, the spectrum becomes appreciably different leading to peculiar thermodynamic effects. This is the essence of the process called size-invariant shape transformation, which give rise to QShE in the physical properties of the system consisting of particles confined within the blue region in between inner and outer domains. 

Here it should also be mentioned that there exist domains, isospectral domains, with distinct shapes having the same sizes but giving rise to exactly the same spectrum \cite{RevModPhys.82.2213}. Naturally these systems won't exhibit any QShE because their energy spectrum are identical. One of the nicest things about the size-invariant shape transformation is that it makes possible to transform domains from one form to another by continuous boundary deformations while still preserving the size-invariance and leading to different spectrum.

One of the most useful aspects of this type of transformation is that it allows to parametrize the QShE and introduce new geometric control variables on the thermodynamic state functions and transport properties of the system \cite{aydin7,aydin11,aydin12}. By this way, one can easily investigate the effects of pure shape dependence on the physical properties of the system.

In Fig. 4(b), we plot the quantum thermal density profiles of the particles confined in the considered domains. When the original and the transformed domains are compared, it can clearly be seen that the density profiles of particles differ radically after making size-invariant shape transformations (either via rotation or translation) on each domain. Due to their wave nature, particles favor to occupy the regions that are far from the impenetrable boundaries of both inner and outer domains. The equilibrium density distribution carefully follows the characteristic shape of the domain before and after any quasistatic change (it is called as adiabatic in the quantum-mechanical context but we will use the term quasistatic as it is used in thermodynamic terminology). More importantly, QBLs of inner and outer domain boundaries start to overlap when they come close to each other. This overlap actually carries information about the shape transformation and they will be useful to understand QShE physically and characterize them analytically, as we will discuss in the next section. 

\subsection{The simplest system exhibiting the quantum shape effect: Particle in a box with a moving partition}

In the previous subsection, we examine how QShE can arise via a size-invariant shape transformation for particles confined in 2D nested domains, which can straightforwardly be extended into 3D. In order to understand and investigate the fundamentals of this effect, we focus on the simplest system that can exhibit the effect. The simplest possible size-invariant shape transformation occurs in a 1D system via the translation of an inner boundary that is imposed. Let's consider a quantum particle in a 1D box that is separated by an impenetrable and infinitesimally thin partition (p) at the center, see Fig. 5(a). Here, the partition (blue dot) in 1D exactly corresponds to the inner domain that we discussed in the case of 2D domains, Fig. 4, and likewise the two ends of the 1D box represents the outer domain boundaries (gray dots). The partition could have also been finite in thickness, but we choose the thickness to be zero for simplicity. Complete separation of the domain into two is also idealistic, in practice one can always think a large enough potential barrier to prevent tunneling. 

Despite the partition has no actual thickness, because it acts as a boundary for both left and right sides of the box, QBLs form an effective thickness of $\delta$ on both sides of the partition giving it an effective thickness of $2\delta$. The partition, as an inner boundary, reduces the effective length of the domain by $2\delta$, unlike the outer boundaries which have thickness of $\delta$ each. This is the basic difference between inner and outer boundaries in terms of how they reduce the effective length of the system. Then the effective length of the box becomes $L_{\mli{eff}}=L-4\delta$. 

\begin{figure}[t]
\centering
\includegraphics[width=0.48\textwidth]{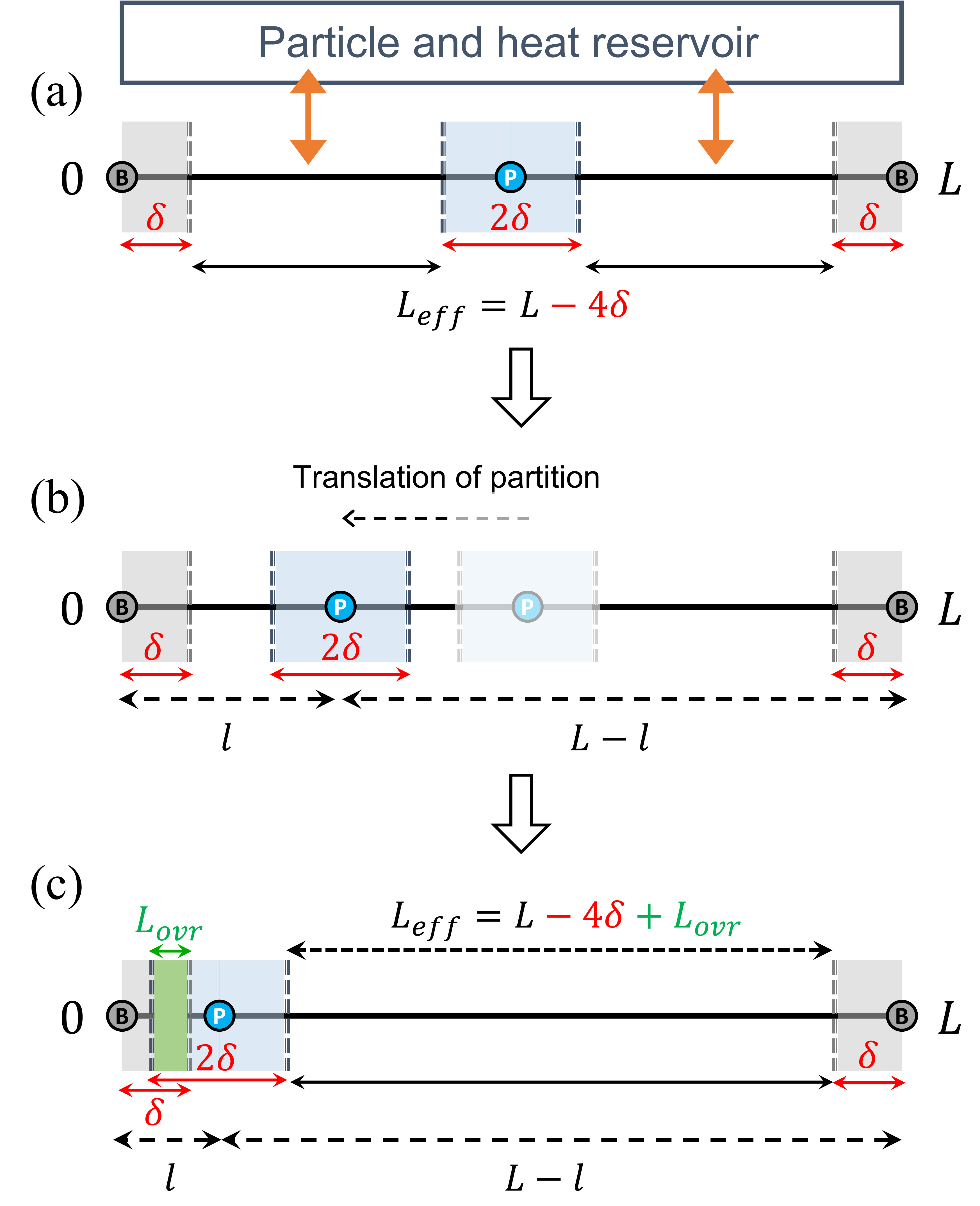}
\caption{Emergence of quantum shape effects in the simplest possible system: quantum particle in a 1D box separated by a movable partition. The system is in contact with a reservoir maintaining the thermal and chemical equilibrium, with fixed temperature $T$ and chemical potential $\mu$, during the movement of the partition. (a) Infinitesimally thin partition divides the box into half, $l=L/2$. Unlike outer boundaries (the two edges), inner boundary (the partition) effectively evacuates double amount of QBLs and so reduces the effective length two times more than the usual outer boundary does. (b) Moving the partition isothermally to left or right (changing $l$) does not change the effective length as long as QBLs of inner and outer boundaries do not overlap with each other. (c) Once QBLs of inner and outer boundaries overlap ($l<2\delta$), QShE start to become appreciable. Effective length is no longer composed only of QBLs, but also of the overlaps of QBLs (denoted by $L_{\mli{ovr}}$) which actually has a positive contribution to the effective length because an effective region equal in amount to $L_{\mli{ovr}}$ is effectively created. Since sizes do not change during the translation of the partition, effective length changes due to shape effect.}
\label{fig:pic5}
\end{figure}

Now let's contact the system with the particle and heat reservoir (maintaining both thermal and chemical equilibrium), and consider moving the partition to either direction (left in the figure) by changing its distance $l$ from the left wall, Fig. 5(b). During such a movement, the domain sizes ($\mathcal{P}=L$ and $\mathcal{N_V}=3$) remains the same at all times, indicating the variation could only be attributed to the shape variation. In other words, the shape that particles are confined in both sides of the box clearly changes from Fig. 5(a) to Fig. 5(b). Although this operation changes the shape of the domain, the resulting QShE can be exponentially small when the partition is far away from the boundaries. In order to have an appreciable QShE, partition needs to come close enough to the domain boundaries so that their QBLs start to overlap with each other, Fig. 5(c). Note that the actual overlap starts earlier than $2\delta$ distance, but as a zeroth order approximation we consider the overlaps closer than $2\delta$ \cite{aydin7}. Only in such a case the effective length of the domain changes with changing $l$. When the distance between the inner (blue dot) and outer (gray dot) boundaries becomes less than $2\delta$, their QBLs start to overlap with each other. QBLs are the regions where particles are practically evacuated due to their wave nature (completely evacuated in zeroth order approach). When QBLs overlap, the effective length of the domain increases because the total amount of evacuated region decreases with the overlap. In other words, size of the free domain that is available for particle occupation increases with the overlap. This is genuinely a QShE, and it cannot be explained by quantum size effects.

A 2D analogue of the case with no overlap between QBLs would be nested square domains with the inner square that is much smaller than the outer square. The rotation of the inner square would still be considered as a size-invariant shape transformation and therefore a shape effect, but QShE won't emerge, if the inner and outer boundaries are not close enough to each other. So the principal condition for QShE to emerge in the first place is already having a strong confinement so that QBLs of inner and outer boundaries can overlap and thereby creating an appreciable difference in the effective sizes.

Note that although we make our analysis in one-dimension, the same problem could very well be designed in 3D or 2D by keeping the other directions constant at certain size. Partition function will just be multiplied with the contributions of the other directions, ought to the fact that Boltzmann factor preserves orthogonality. In fact, we will calculate thermodynamic properties by considering the same problem in 3D, in Sec. V. However, it is convenient to simply study the 1D problem in this section, to investigate the characteristics of the QShE without loss of generality.

\subsection{Quantum shape effect is not just a stronger quantum size effect}

At this point, let's turn back to the question that we asked at the end of Sec. II(D): whether QShE can be interpreted as just a strong version of quantum size effects or not. To explicitly compare quantum size and shape effects, we could examine them under the same parameter $l$, but corresponding to a different physical operation on each case. We normalize the lengths in terms of $\delta$ for generality.


In a box with length $l$ without any partition, changing the length of the box corresponds to the size effect, Fig. 6(a), which is parametrized by two size variables $\mathcal{P}=l$ and $\mathcal{N_V}=2$. Note that when the box is shrunk to a length $l<2\delta$, QBLs of the boundaries start to overlap. However, this is still a size effect, simply because the operation of changing $l$ only changes the size variables of the domain. When we calculate the partition function for a box with changing length from 0 to $L$, we see that for $L<2\delta$ the partition function effectively becomes zero, see orange curve in Fig. 6(c). This is because the confinement in the system is so extreme that the thermal excitation probabilities are exponentially low. As a matter of fact, the ground state has already much higher energy than the thermal energy $k_BT$. Another point of why this type of overlap cannot be interpreted as a QShE is because they do not lead to an increment of the effective length of the domain, which is a signature of QShE. To give more perspective on this, let's consider the corners of a square domain in 2D, like in Fig. 4(b). Perpendicular sides of the domain create a $\delta^2$ overlap of the QBLs of the outer boundaries. This type of overlap is associated with the number of vertices correction, see Fig. 2.14 in Ref. \cite{aydinphd}. Overlaps of the QBLs of outer boundaries give rise to a lower dimensional QSE contributions, whereas overlaps of the QBLs of outer boundaries with the QBLs of inner boundaries give rise to the QShE contributions. The difference that is shown in Fig. 6(c) is a clear manifestation of this fact.

In the case of QShE, $l$ corresponds to the movement of the inner boundary inside the domain from 0 to $L$, Fig. 6(b). This operation does not change the size variables at all (both the length and number of vertices stay constant). Any change in the partition function is due to QShE, purple curve in Fig. 6(c). The effects of size and shape variation on the partition function converges after $l\gtrsim 2\delta$, because $l$ becomes so close to $L$ that the length of the right side of the box becomes effectively zero and do not contribute to the partition function. In such a case moving the partition to the right only corresponds to the extension of the left part of the box, which basically happens to be a very similar operation to a size effect. Nevertheless, it should be noticed that in the interval where QSE and QShE give similar results, increasing $l$ decreases QSE but increases QShE. Therefore, QSE and QShE do not originate from the same mechanism. The difference around $l=2\delta$ might seem to be a small one, but in fact it is not. Influence of QShE become considerably important in larger dimensions (where many local constrictions can exist) and for other thermodynamic properties derived from the partition function.

To summarize, there are three main mechanisms separating QShE from QSE: (1) QShE occurs under fixed sizes and it is characterized by its own distinctive geometric coupling parameters (in this example it is $l$, the distance of the partition from the left boundary). (2) QSE and QShE have considerably different effects on the eigenspectra. For instance, size effect only scales the spectra whereas shape effect has a much more complicated influence, which deserves to be investigated in another paper in detail. (3) The more QSE the less effective volume, whereas the more QShE the more effective volume. In other words, QShE has the exact opposite effect on the system than QSE: the more QShE the system exhibit, the more deconfined the system is. While QSE contributes to the confinement of the domain, QShE contributes to its deconfinement. These three points show that QShE are fundamentally different than QSE. Nevertheless, QShE is, in a sense, an "additional" effect on QSE, so QSE is a prerequisite for the observable existence of QShE. Both effects are single-particle and statistical-mechanical effects caused by energy quantization due to quantum confinement. In order for QShE to show up, the domain sizes should be small enough so that inner and outer boundaries can become close to each other around a few thermal wavelengths. Therefore, QSE is the result of a global (overall) confinement, whereas QShE is related to the local confinement (narrow parts, constrictions etc.), which effectively leads to the global deconfinement of the domain.

\begin{figure}[t]
\centering
\includegraphics[width=0.48\textwidth]{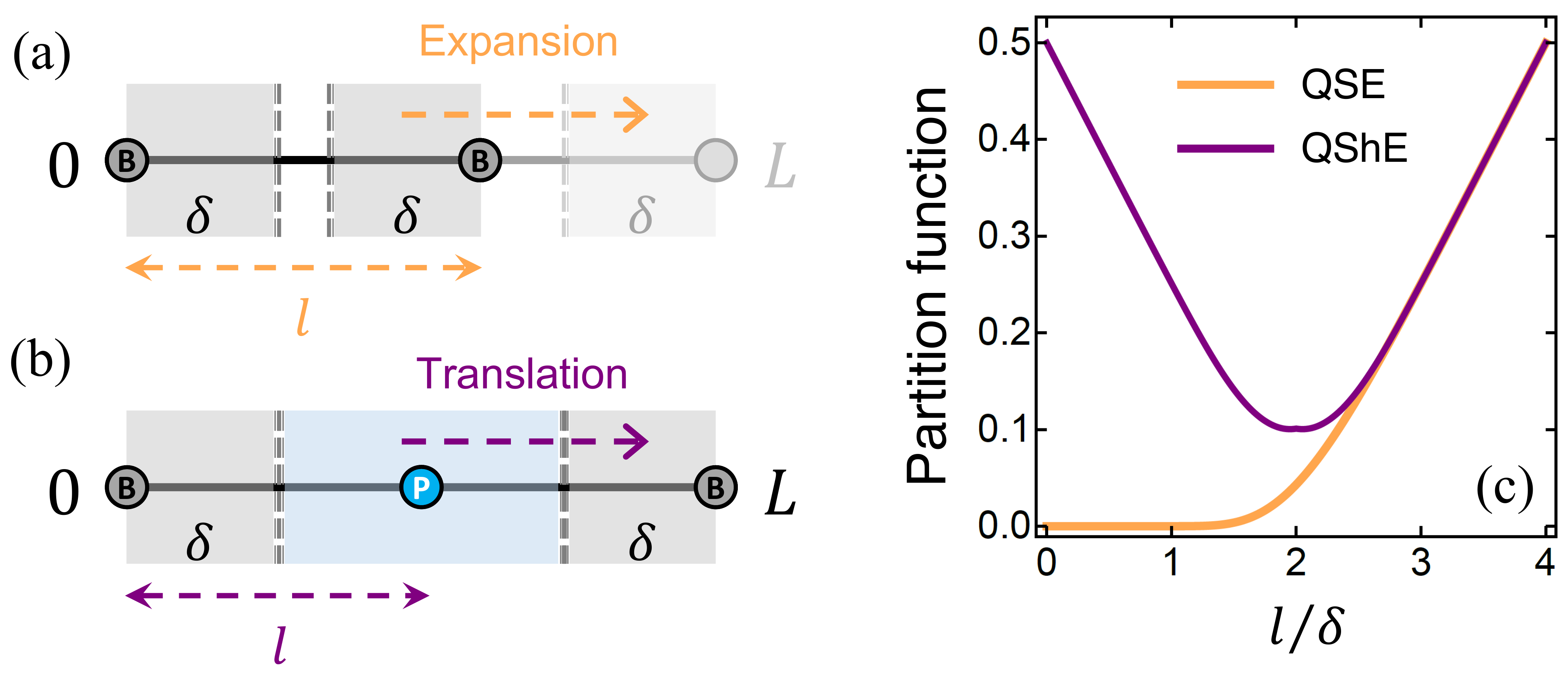}
\caption{Difference between quantum size and shape effects. (a) Demonstration of extreme quantum size effects in particle in a box with changing length $l$ with the range $0\leq l\leq L$. (b) Quantum shape effects in particle in a box with fixed length $L$ but moving partition with $0\leq l\leq L$. (c) Partition function changing with $l$. Quantum size and shape effects give almost the same result after box length becomes larger than $2\delta$. Around and below $2\delta$, they considerably differ. Below $2\delta$, there is effectively no space left for particles to occupy within the box in QSE case (a), whereas quantum shape effects increase the effective length and partition function increases in QShE case (b). All processes are done under thermal and chemical equilibrium with the reservoir.}
\label{fig:pic6}
\end{figure}

\section{Analytical approximations for calculating quantum shape effects}

We seek for analytical methods to predict QShE, to understand the underlying physics behind it as well as to get rid of cumbersome numerical calculations.

1D partition function for a single particle obeying Maxwell-Boltzmann statistics is written in its exact summation form
\begin{subequations}
\begin{align}
\zeta(L) &=\sum_{i=1}^\infty\exp \left[-\frac{\pi}{4}\left(\frac{\lambda_{th}}{L}i\right)^2 \right] \\
&=\frac{1}{2}\vartheta_3\left[0,\exp\left(-\alpha(L)^2\right)\right]-\frac{1}{2}.
\end{align}
\end{subequations}
where $\alpha(L)=\sqrt{\pi}\lambda_{th}/(2L)$ is defined as the confinement parameter. As is shown in Eq. (9b), the partition function can also be analytically expressed by the elliptic theta function of third kind, $\vartheta_3$. We deliberately emphasized that $\zeta$ is a function of $L$ (it is also a function of temperature via $\lambda_{th}(T)$). Now that we have a domain separated into two by a partition, we can write the partition function for such a domain simply by separately calculating the partition function for both parts of the box and add them together. Since the partition is impenetrable, two parts of the domain can be thought as two distinct domains having length $l$ and $L-l$, geometrically coupled to each other (by design) via the constraint that their addition should give $L$. In fact, union of their separate eigenspectra exactly gives the eigenspectrum of the whole domain, i.e. $E_i(L,l)=E_{i^{\prime}}(l)\cup E_{i^{\prime}}(L-l)$. Then, the partition function for the whole system can be written as
\begin{equation}
\begin{split}
Z(L,l)=\zeta(l)+\zeta(L-l).
\end{split}
\end{equation}
We shall denote this combined partition function with the letter $Z$ in order to distinguish it from the partition function $\zeta$ for each side of the box. We stress that no approximation has been done in writing of this expression, i.e. exact.

The exact effective length can then be calculated directly using the exact partition function from Eq. 8(b) as follows
\begin{equation}
\begin{split}
L_{\mli{eff}}=Z(L,l)\lambda_{th}(T).
\end{split}
\end{equation}

Now we would like to analytically express the partition function under QShEs. To be able to do that we need to quantify the overlaps of QBLs when they come close to each other. Overlaps of QBLs can be interpreted and calculated in several ways within the QBL approach. Here, we would like to give the most useful one which is the overlapped QBL approach introduced in Ref. \cite{aydin7}. Next, we will apply a new approach based on dimensional transitions, the first time in this paper. Both approximations generate similar results by different analytical expressions, but more importantly, they provide physical understanding of QShE from different perspectives.

\subsection{Overlapped QBL approximation}
We invoke QBL method to construct an approximation for obtaining shape dependent thermodynamic quantities. Overlapped QBL approximation is first introduced in Ref. \cite{aydin7} to analytically predict and physically interpret QShE. The approximation is illustrated in Fig. 5(c) where QBLs of inner (partition) and outer boundaries overlap when the partition is closer than $2\delta$ from the boundary. Normally effective length is calculated by removing the QBLs from the actual domain size, i.e. $L_{\mli{eff}}^0=L-L_{\mli{qbl}}$. However, in the case where QBLs of inner and outer boundaries overlap, the amount of overlapped length has become oversubtracted. Overlapped QBL approach approximates the effective length by adding the excess removal of the overlap length. Then, effective length in overlapped QBL approximation is written as
\begin{equation}
\begin{split}
L_{\mli{eff}}^\mli{ovr}=L-L_{\mli{qbl}}+L_{\mli{ovr}},
\end{split}
\end{equation}
where $L_{\mli{qbl}}=4\delta$ is the length of the total QBLs inside the domain and $L_{\mli{ovr}}$ term quantifies the amount of length that is constituted by the overlaps of inner and outer QBLs. Within the overlapped QBL approximation, overlap length can be calculated by the following piecewise function
\begin{equation}
L_{\mli{ovr}}=
\begin{cases}
2\delta-l & l\leq 2\delta \\
2\delta-(L-l) & L-l\leq 2\delta \\
0 & 2\delta<l<L-2\delta
\end{cases}
\end{equation}

When $l<\delta$, left QBL of the partition outflows from the left boundary. However, the overlap and outflow of QBLs have the same effect on the effective length, so we treat both under the same footing and do not distinguish the outflow contribution from the overlap contribution to the effective length. Then, within the overlapped QBL approximation, the total single particle partition function can be analytically written as
\begin{equation}
\begin{split}
Z_{\mli{ovr}}=\frac{L_{\mli{eff}}^\mli{ovr}}{\lambda_{th}}=\frac{L_{\mli{eff}}^\mli{ovr}}{4\delta},
\end{split}
\end{equation}
which is functions of temperature $T$, size $L$, and shape $l$.

It has been shown previously that overlapped QBL approximation is successful in predicting quantum shape dependent properties and capturing their functional behaviors. In addition to that, it provides a physical explanation for the increase in effective sizes when QShE are prominent, because overlap regions contribute positively to the effective size of the domain. It also transforms a complicated quantum mechanical problem into a simple geometrical calculation.

\subsection{Dimensional transition approximation}

Another way of interpretation and prediction of QShE can be done by introducing the dimensional transition approximation. This time, rather than introducing the concept of effective length, we focus on the dimensional change in the partition function due to the movement of partition. QSE and QShE can trigger dimensional changes in the representation of the system's properties. For example, due to strong confinements in one or more directions, it becomes possible to represent the physical properties of the particles confined in these systems via their lower dimensional expressions, which is why they are usually called lower dimensional materials. We can take advantage of these dimensional changes to calculate the QShE even more precisely. When the partition moves from center to left, left domain becomes more and more confined so that the partition function transitions from 1D to 0D representation. Under such extremely strong confinements ground state takes over and 0D representation becomes sufficient \cite{aydin2,PhysRevE.104.054110}. In fact, as it has been shown in Ref. \cite{aydin2}, this takeover occurs at a specific point in confinement space, approximately at $L=L^{*}\approx 0.7\lambda_{th}$. For comparison, this transition occurs approximately at $2.8\delta$, which means that the dimensional transition approximation can predict QShE behaviors more precisely, even before the overlaps begin.

\begin{figure}[t]
\centering
\includegraphics[width=0.48\textwidth]{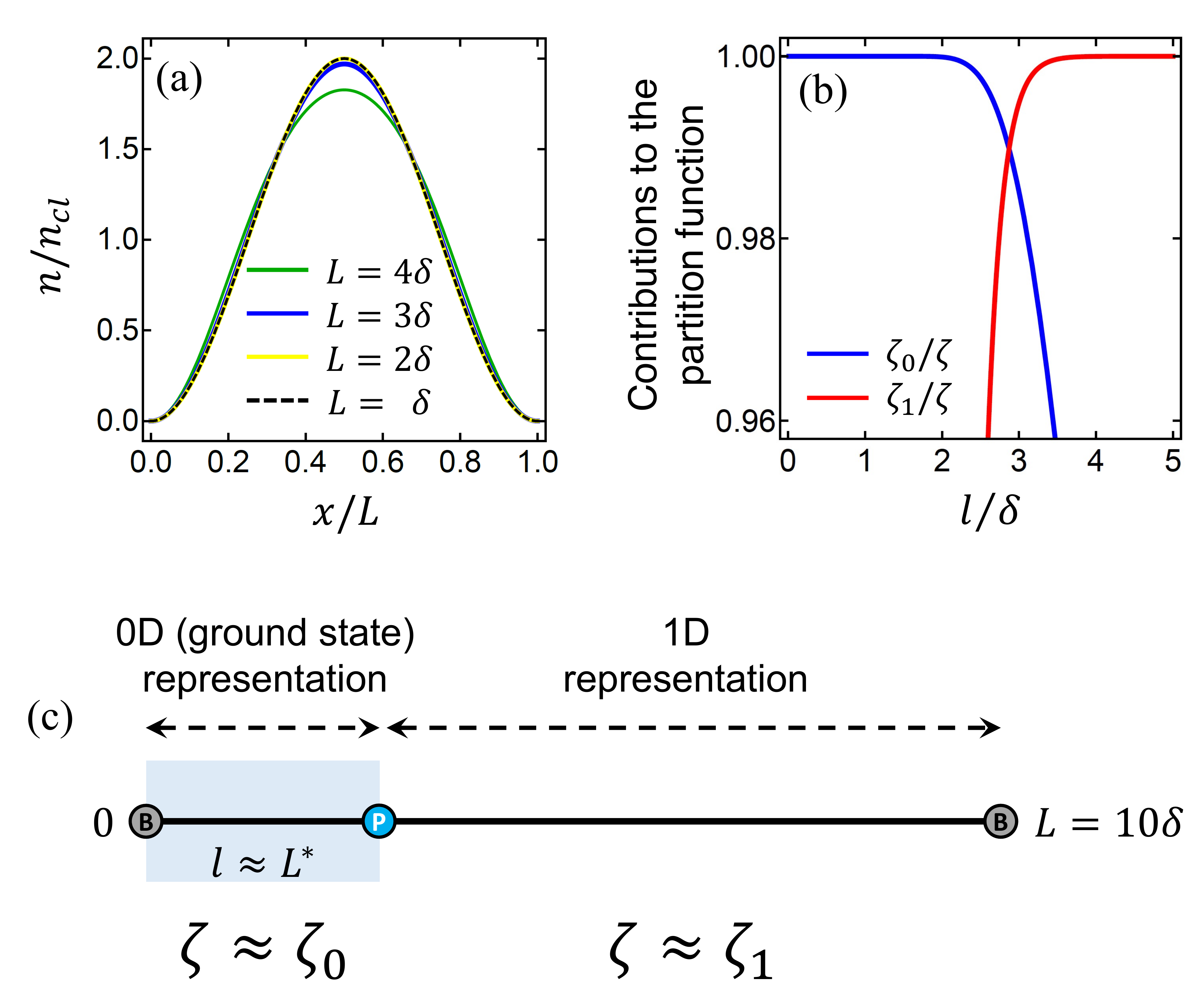}
\caption{(a) Saturation of quantum thermal density towards the ground state density with reducing size. Dimensionless quantum thermal density distribution of a particle in a box for different box lengths measured by the QBL thickness $\delta$. For $L\lesssim L^{*}$ ground state density takes over which justifies the dimensional transition approximation. (b) Comparison of the contributions of 0D and 1D partition function to the exact one. Ratios of $\zeta_0/\zeta$ and $\zeta_1/\zeta$ go to unity when $l\lesssim L^{*}$ and $l\gtrsim L^{*}$ respectively. (c) Illustration of the domain when partition is at $l\approx L^{*}$ in which case, partition functions of the left and right parts of the domain can be described by zero- and one-dimensional partition functions respectively.}
\label{fig:pic7}
\end{figure}

In Fig. 7(a) we calculate quantum thermal density for the left part of the domain with $l$ changing from $4\delta$ to $\delta$. Quantum thermal density (Eq. 7) is normalized by the classical density $n_{cl}$. It is seen that quantum thermal density approaches to the ground state density after around $L^{*}\approx 0.7\lambda_{th}$. This means for the left side of the box, it is enough to consider only the ground state when $l<L^{*}$. The principal quantity is the partition function, from which we can derive all other thermodynamic properties. Therefore we can offer the following piecewise function for the dimensional transition of partition function of the left side of the box:
\begin{equation}
\zeta_{\mli{DTA}}(L^\prime)=
\begin{cases}
\zeta_0(L^\prime)=\exp\left(-\frac{\pi}{4}\frac{\lambda_{th}^2}{{L^{\prime}}^2}\right) & L^\prime\leq L^{*} \\
\zeta_1(L^\prime)=\frac{L^\prime}{\lambda_{th}}-\frac{1}{2} & L^{\prime}> L^{*}
\end{cases}
\end{equation}
where we replace $l$ with $L^\prime$ to indicate that it is now a general parameter which can then be replaced by the proper lengths in the calculation so that partition function for either side of the box can be defined by Eq. (15). For $L^\prime\leq L^{*}$ (0D representation) partition function approaches to its ground state value ($i=1$) at Eq. (9a). For $L^\prime> L^{*}$ (1D representation), it is represented by Eq. (4).

The accuracy of Eq. (15) can be seen in Fig. 7(b) where the ratios of the contributions of 0D and 1D representations to the the partition function. 0D and 1D representations quite accurate when $l<L^{*}$ and $l>L^{*}$ respectively. Fig. 7(c) shows how different parts of the box is treated within different dimensional representations. Using Eq. (15), total partition function of the whole system can be calculated as usual $Z_{\mli{DTA}}(L,l)=\zeta_{\mli{DTA}}(l)+\zeta_{\mli{DTA}}(L-l)$, which explicitly shows both parts of the box treated within their own dimensional representations.

\begin{figure}[t]
\centering
\includegraphics[width=0.48\textwidth]{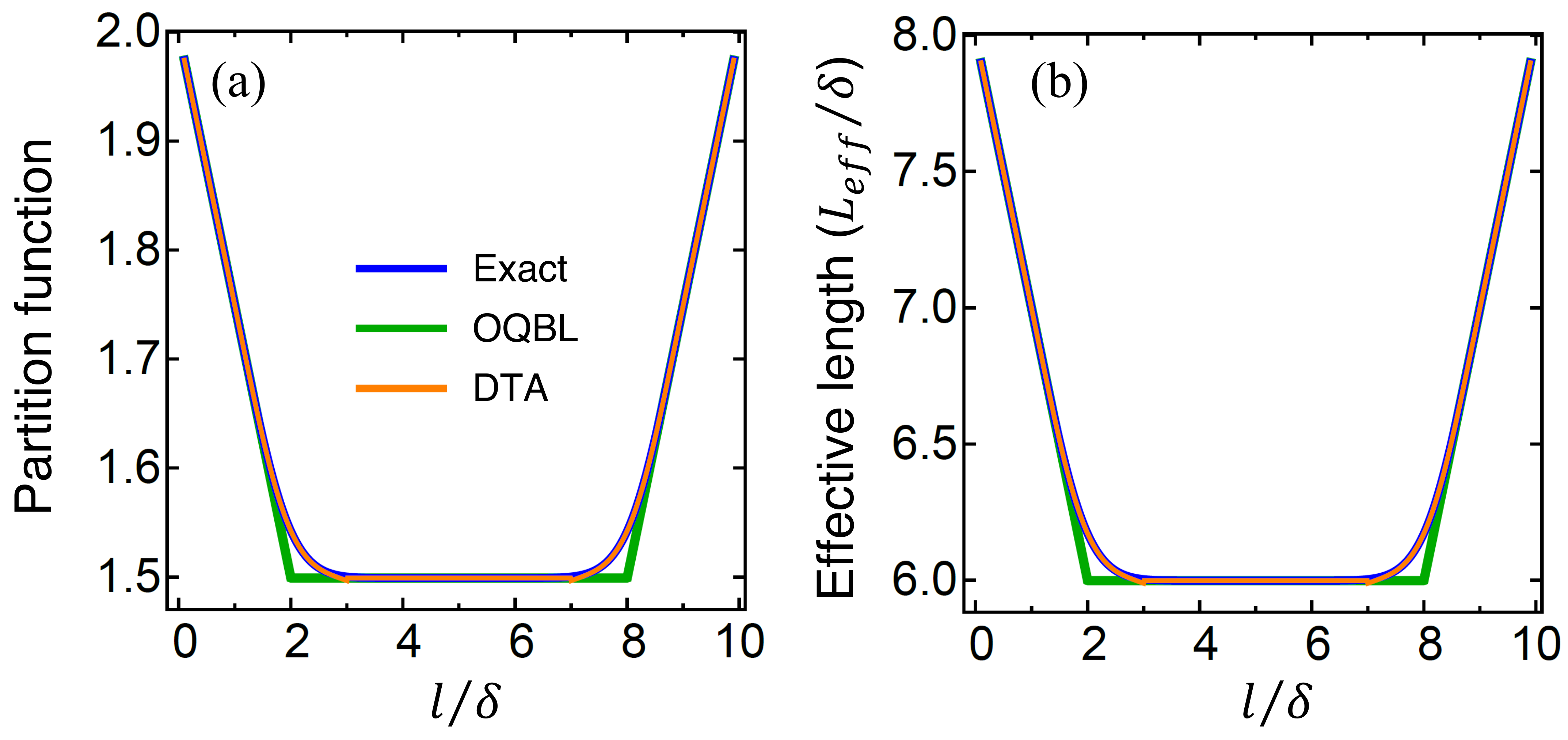}
\caption{Comparison of the accuracies of overlapped QBL (OQBL) and dimensional transition approximations (DTA) with the results of exact summation. (a) Partition function and (b) effective length changing with the partition position that is moved from one end to the other end of the box.}
\label{fig:pic8}
\end{figure}

In Fig. 8, we give comparisons of overlapped QBL and dimensional transition approaches with the exact numerical calculations in terms of partition function and effective length changing with $l$. The functional behaviors of partition function and effective length are the same as they are directly proportional to each other via Eq. (8) with the proportionality factor of $1/\lambda_{th}=1/(4\delta)$. From the perspective of overlapped QBL, the overlaps of QBLs starts when the distance between the box boundary and the partition is less than $2\delta$, producing the increase in effective length. The lowest values (plateau) of effective length is $L-4\delta$ where $2\delta$ is coming from QBLs of left and right boundaries of the box and the other $2\delta$ is coming from the QBLs of the partition which is formed left and right sides of it. The maximum values of the effective length occurs when the partition is exactly at the domain boundaries which causes the left QBL of the partition to completely vanish (since it outflows), giving the effective length $L-2\delta$, which is basically the effective length of a box without partition. Even though dimensional transition approximation does not rely on defining effective length, we can calculate the corresponding effective length within that approximation via $L_{\mli{eff}}^{\mli{DTA}}=Z_{\mli{DTA}}(L,l)4\delta$ to compare with the overlapped QBL one. Fig. 8 shows that dimensional transition approximation gives quite accurate results for all values of $l$ with a negligible error around the transition point $L^{*}$. The effective length (or volume in 3D) concept will be central for the explanation of the behaviors of thermodynamic properties in the next section. Since the errors of dimensional transition approximation are quite negligible, the difference won't be distinguishable in the figures. Nevertheless, we will use the exact forms (based on summations over eigenvalues) of thermodynamic quantities henceforward.

\section{Thermodynamic properties under quantum shape effect}

In this section, we calculate the thermodynamic properties of non-interacting particles confined in a domain exhibiting the QShE. Unlike in previous sections, this time we consider a 3D box for the sake of including many particles. Two directions ($L_x$ and $L_y$) are chosen to be large and one direction ($L_z$) is extremely confined. In particular, we choose $L_x$ and $L_y$ to be $1000\delta$ and $L_z$ to be $10\delta$ as above. The partition is inserted in parallel to the macro directions and perpendicular to the nano direction.

For $N$ non-interacting particles, the total partition function is written as $\mathcal{Z}=Z_{3D}^N/N!$ where the single particle partition function of the composite system is
\begin{equation}
\begin{split}
Z_{3D}=\zeta_{3D}(L_x,L_y,L_z=l)+\zeta_{3D}(L_x,L_y,L_z=L-l),
\end{split}
\end{equation}
and the single particle partition function for the 3D box is
\begin{equation}
\begin{split}
\zeta_{3D}=\left(\frac{L_x}{\lambda_{th}}-\frac{1}{2}\right)\left(\frac{L_y}{\lambda_{th}}-\frac{1}{2}\right)\zeta(L_z).
\end{split}
\end{equation}
Note that in MB statistics we are able to write the 3D partition function as the separate products of partition functions of each direction. For increased accuracy, we also consider the QSE corrections for $L_x$ and $L_y$ directions as well. Total number of particles inside the box can be written as $N=e^{\mu/(k_BT)}Z_{3D}$. Number of particles in left and right compartments of the box is written as $N_L=e^{\mu/(k_BT)}\zeta_{3D}(L_x,L_y,l)$ and $N_R=e^{\mu/(k_BT)}\zeta_{3D}(L_x,L_y,L-l)$ respectively. Chemical potential $\mu$ can be straightforwardly calculated from $N$. Now, using the total $N$-particle partition function $\mathcal{Z}$, we calculate Helmholtz free energy, entropy and internal energy respectively as $F=-k_BT\ln \mathcal{Z}$, $S=-\partial F/\partial T$, $U=F+TS$.

\begin{figure*}[t]
\centering
\includegraphics[width=0.98\textwidth]{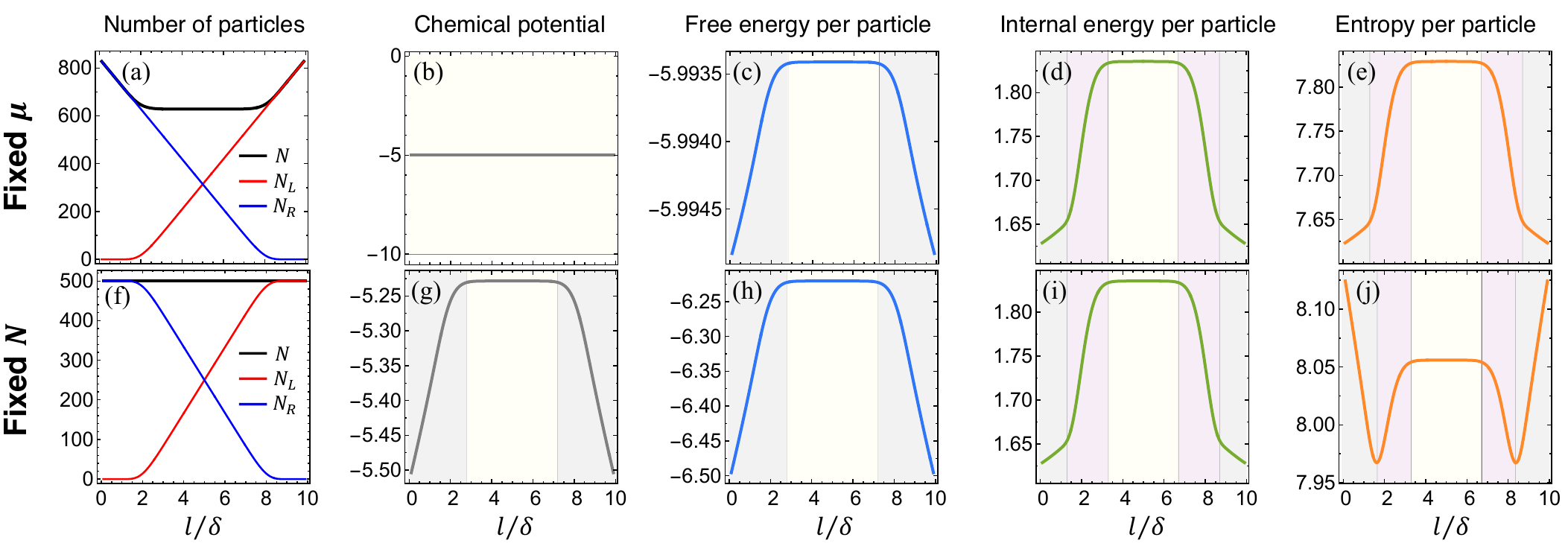}
\caption{Thermodynamic properties varying with the shape variable $l$. (Top row) The system is in thermal and chemical equilibrium with the heat bath so that both the temperature $T$ and the chemical potential $\mu$ are constant during the process. (Bottom row) The system is in thermal equilibrium while total number of particles $N$ is constant. Particles are allowed to transfer between compartments through a permeable partition. Variation of number of particles in the left and right compartments of the box ($N_L$ and $N_R$) is shown by red and blue curves respectively. Background color codes in the plots of normalized thermodynamic properties are chosen as follows: Yellow: unchanged region (no quantum shape effect), Gray: ordinary behaviors (similar to the quantum size effect), Pink: peculiar behaviors (special to the quantum shape effect).}
\label{fig:pic9}
\end{figure*}

We investigate how thermodynamic properties change when the partition is varied. We consider two cases for thermodynamic analysis: fixed chemical potential and fixed number of particles. In the first case, the box is in chemical equilibrium with the reservoir where we allow particle exchange to keep the chemical potential constant. Particles can also transfer between the compartments over the reservoir. We set $\mu/(k_BT)=-5$, ensuring the applicability of Maxwell-Boltzmann statistics. In Fig. 9(a), we show the variation of particle number with partition position $l$ (normalized to $\delta$). Total particle number reduces when the partition moves away from boundaries. This is correlated with the behavior of effective volume (see Fig. 8b). When the partition is moved from the left boundary to the center, effective volume decreases, making the system effectively more confined and causing particles to escape into the reservoir. The smaller the effective volume, the less particles there are inside the box. When the partition is up to $2\delta$ away from boundaries, the particle number in the smaller compartment is effectively zero. Because of the imbalance between the lengths of the compartments and extremely high confinement of the smaller compartment, all particles prefer to occupy the larger compartment or to the reservoir. Chemical potential, normalized by $k_BT$, is constant, Fig. 9(b).

Transitions into thermodynamically more stable states are dictated by the variations in Helmholtz free energy. Behavior of normalized Helmholtz free energy per particle ($F/(NkT)$) at fixed $\mu$, varying $N$ is shown in Fig. 9(c). It takes the form of $F/(NkT)=\mu/(k_BT)+(\ln{N!}-N\ln{N})/N$. The increase in the total number of particles causes a slight decrease (noticeable only at the third decimal) in the free energy per particle when the partition is near the boundaries, denoted by the gray background color indicating the ordinary/expected behaviors. Since the chemical potential stays constant, the change in free energy per particle is solely due to the change in total number of particles during the movement of the partition. Essentially, it is a consequence of the classical indistinguishability of particles, since the indistinguishability correction in free energy brings additional dependence on the total number of particles inside the system. Hence, using the Stirling approximation is not appropriate in this case, as it would give $F/(NkT)=\mu/(k_BT)-1=-6$ which fails to capture the small variation in free energy per particle. Free energy stays almost constant when the partition is more than around $L^{*}$ away from the boundaries of the domain, denoted by the yellow background color indicating the unchanged region. The background color codings apply to all other subfigures in Fig. 9.  The unchanged regions having no meaningful QShE can be explained via two different perspectives: (1) There is no overlap of QBLs until the distance between the partition and the boundary is less than $2\delta$ (overlapped QBL perspective). (2) Partition function of the smaller domain consists only of the contribution of the ground state after around $L^{*}$ distance (dimensional transition perspective). Substantial change starts to occur when the partition is closer to the either side of the domain than around $L^{*}$. Since figures are perfectly symmetric around the center $L/2=5\delta$ in x-axis, we shall focus on the left side of the figures and interpret the behaviors of thermodynamic quantities considering the partition moving from center to the left.

Variation of normalized internal energy per particle ($U/(NkT)$) at fixed $\mu$, varying $N$ is shown in Fig. 9(d). It takes the approximate form of $U/(NkT)\approx[u(l)+u(L-l)]/Z(L,l)+1$, where $u(L^{\prime})=\sum_{i=1}^{\infty}[\alpha(L^{\prime})i]^2\exp[-[\alpha(L^{\prime})i]^2]$. Classically, the normalized internal energy per particle is $U_{cl}/(NkT)=3/2$. Due to quantum confinement, internal energy is above $3/2$ during the whole process of varying $l$. It is reduced near the boundaries due to the existence of QShE which decreases confinement energy contribution by increasing the effective volume of the domain. The more interesting part is the noticeable change of the slope of internal energy per particle. We labeled the sharper drop as the peculiar behavior and the other one as the ordinary behavior, for the reasons that will become clearer below. The initial sharp decrease in the peculiar region (denoted by the pink background color) occurs because of the increase in effective volume which makes the system effectively less confined and reduces the confinement energy. Since the rate of change in effective volume is larger in the peculiar region due to overlaps (e.g. compare it with Fig. 8b), reduction in internal energy is faster in this region. In the ordinary region, on the other hand, the slope changes because of the fact that when the partition gets closer to the boundary, contribution of the overlaps only consists of the expansion of the larger compartment, as the region between the boundary and the partition has already been evacuated in the smaller part. There are almost no particles left in the smaller compartment of the box and the system effectively turns into the expansion of the larger compartment. Due to this reason, the functional behaviors of thermodynamic properties in this region are ordinary (in the sense that one expects from the isothermal expansion/compression of a confined gas under QSE). The behavior of internal energy per particle is completely independent of the variation in the total number of particles.

One of the most interesting consequences of QShE is seen in the peculiar behavior of entropy. In the classical thermodynamics of gases, free energy and entropy behaves oppositely with respect to changes in volume in an isothermal expansion/compression process. On the other hand, this is not always the case for systems exhibiting QShE \cite{aydin7}. Variation of normalized entropy per particle ($S/(Nk)$) at fixed $\mu$, varying $N$ is shown in Fig. 9(e). Due to a negligible change in free energy, entropy mimics the behavior of internal energy. Therefore, the decreases in entropy per particle are solely due to the decreases in the confinement energy per particle. To understand this behavior from the phenomenological thermodynamics perspective, consider the entropy per particle of the left and right compartments separately. During the variation of the partition, the entropy of the smaller compartment decreases, while that of the larger one increases in accordance with the expectations. Total entropy decreases both in the peculiar and ordinary regions. Because in the peculiar region, the decrease in entropy of the smaller compartment dominates the total change in entropy. In the ordinary region, on the other hand, the slope changes because the total number of particles increases as there are effectively no particles in the smaller compartment, whereas the particle number increases in the larger one.

Now we proceed with our next analysis: keeping the total number of particles fixed in the system by allowing particle exchange through compartments via the permeable partition, but not allowing particle exchange with the reservoir. Note that despite the permeability, Dirichlet condition can effectively be satisfied by creating tiny holes smaller than $\lambda_{th}$ on the partition wall. In this way, when confinement increases the energy levels, particles with smaller wavelengths can penetrate through the holes to the other compartment. In this case, the reservoir acts only as a heat bath, keeping the temperature fixed in the whole system. Particle numbers in the left and right compartments behave similar to the previous case, Fig. 9(f), except this time they cannot escape into the reservoir and all particles accumulate to the larger compartment when the partition is too close to the boundaries. Fixing the total number of particles causes chemical potential of the system to vary during the process, Fig. 9(g). Chemical potential decreases because increase in effective volume causes effective density to decrease. The decrease in chemical potential near the boundaries is also correlated with the increase in the total number of particles in the previous case, e.g. compare with Fig. 9(a). Note that the chemical potentials of left and right compartments are equal to each other.

In Fig. 9(h), Helmholtz free energy per particle steadily decreases from $l=L^{*}$ to $l=0$ at constant $T$ and fixed $N$. Free energy tells the direction of the thermodynamic transition under quasistatic process. Thus, when the system is prepared in such a way that the partition is positioned at a distance between $0<l<L^{*}$, it will spontaneously move to the left boundary, assuming no friction and no other forces acting. Occupiable modes in the larger compartment prevail over the ones in the smaller compartment and a quantum force emerges \cite{aydin7}, bringing the inner and outer boundaries closer to each other. In this sense, the quantum force that will act on the partition is quite similar to the Casimir force. The reason of this thermodynamic behavior is directly because of the fact that effective volume of the system increases due to QShE, when the partition moves from $l=L^{*}$ to $l=0$ isothermally. In other words, existence of QShE causes effectively more available domain for particles to occupy, which basically amounts to expansion. In fact, such a spontaneous movement of partition to the boundary is analogous to the isothermal expansion of a confined gas, even though the actual volume of the domain remains unchanged. Here, the expansion due to QShE is an effective one, keeping the actual volume constant and changing the effective volume only. The behavior of internal energy per particle at constant $T$ and fixed $N$, shown in Fig. 9(i), is identical to the previous case. This is because internal energy per particle is independent of the variations in chemical potential as well.

The behavior of entropy per particle is shown in Fig. 9(j). When the partition is moved from $l=L^{*}$ to $l=0$ isothermally, entropy first decreases and then increases, whereas the free energy steadily decreases during the process. There are two regions where entropy behaves differently during a smooth variation of a thermodynamic control variable, $l$. These regions are distinguished by the minimum of entropy, which is a complicated function of $\delta$ and depends strongly on the geometry of the system. In the ordinary region, thermodynamic properties exhibit the usual behaviors as explained above. Similar to isothermal expansion of a gas, entropy increases while free energy and internal energy are decreasing. The behavior of entropy in the peculiar region is unseen in classical thermodynamics. This peculiar behavior can be explained via two different perspectives. The first perspective is associated with the temperature sensitivity of the effective volume which comes from the temperature dependence of QBLs \cite{aydin7}. As it has been investigated in Ref. \cite{aydinphd}, there are two terms determining the entropy behavior under QShE. Depending on the competition between these two terms, entropy could either increase or decrease with respect to changes in the shape of the system. Although the entropy decreases in the direction of spontaneous transition (dictated by free energy minimization), this does not violate the second law, because the total entropy of the system and the bath stays constant in a reversible process. So the system essentially exchanges heat with the bath to keep the temperature constant. The second perspective is related with the distinct characteristics of the eigenspectra under QShE and their influences in the partition function via thermal probabilities. Normally, the partition function linearly depends on the actual volume of the system. However, in the peculiar region due to QShE, the partition function exponentially increases (decays), see Fig. 8(a), when the partition moves away from (closer to) the boundary. The same behaviors are mimicked by the effective volume as well. Exponential increase in the partition function and effective volume occur whenever there is a sharp increase in the thermal occupation of ground and low-lying states compared to the other states in the spectrum. Because the ground state makes the biggest contribution to the partition function. Our spectral analysis shows that QShE causes a nonuniform change in the eigenvalues (see Figs 3.5 and 3.6 in Ref. \cite{aydinphd}) and increase in the thermal occupation probability of the ground state in the peculiar region when partition moves towards the boundary. Increased ground state occupation results partition function to exponentially increase, as its first term, the ground state, dominates its behavior. Entropy of the system decreases exactly due to the increase in the ground state occupation probability, which becomes dominant in the entropy. We investigate the consequences of the spectral features of QShE in another paper in detail \cite{specsist}. Eventually, this is basically a competition of two different mechanisms determining the behavior of entropy (uniform and non-uniform scaling in eigenvalues) and which one prevails depends on the exact geometrical configuration of the system.

\section{Discussion and Conclusion}

In this paper, starting by revisiting quantum size effects, we showed the origin of a distinct physical phenomenon called the quantum shape effect appearing at the nanoscale. We considered the simplest system exhibiting the characteristic properties of the effect, namely non-interacting particles in a box with moving partition under the quasistatic process. We demonstrated how quantum size and shape effects are different from each other and how they are similar in some aspects. Furthermore, we applied a new analytical method based on the dimensional transition of partition function \cite{aydin2} and accurately predicted the QShE. Finally, we investigated the changes in thermodynamic properties due to the QShE under various equilibrium conditions. We find that thermodynamic properties, especially the entropy per particle, exhibit peculiar behaviors that are unseen in classical thermodynamics.

Originally, QShE are introduced in a core-shell quantum wire \cite{aydin7} and characterized by the rotation angle of the core wire. It was due to a rotational size-invariant shape transformation. Here, we characterized it by the position $l$ of the partition in the box, originating due to a translational size-invariant shape transformation. From a different perspective, QShE can appear in the case of strongly confined multiple systems (double here) geometrically coupled via the parameter $l$ (here). In that sense, QShE (changing the position of the partition $l$) actually mediates the coupling between two "separate" boxes. If you move the partition to left, for instance, left part of the box will contract and right part of the box will extend at the same amount. QShE is basically an effect caused by the inner and outer boundaries of a domain getting substantially close to each other. Other less confined parts of the domain are also affected by this congestion, creating a global coupling over the relevant shape parameter. Spectrum of the system is affected in a unique fashion by the QShE, which is the deeper cause of the observed peculiar behaviors in thermodynamic quantities. We will investigate the spectral characteristics of QShE in detail in a separate work. Furthermore, this emerged coupling also implies that QShE cannot be explained by any form of QSE as QSE cannot reproduce the effects of QShE generated by these additional shape parameters. One could also insert more than one partition to create additional shape parameters and more complicated couplings.

Despite the fact that nature of this so-called coupling is classical, its consequences at nanoscale are quantum mechanical due to the energy quantization via quantum confinement. In fact, constructing quantum thermal machines by taking advantage of the discrete energy spectrum has become quite popular in the last decade \cite{PhysRevE.72.056110,Uzdin_2014,Campisi2016,PhysRevE.93.050203,aydin7,PhysRevLett.120.170601,PhysRevA.99.022129,e23050536,PhysRevE.104.044133}. In a nutshell, both quantum size and shape effects are inherently quantum effects as they are direct consequences of the prominence of the discrete spectrum and the wave nature of particles. Also from the QBL perspective, when Planck's constant goes to zero, QBL disappears so that both quantum size and shape effects vanish (i.e. particles would occupy the space homogeneously in Fig. 3). 

Besides fundamental importance, QShE provides a novel way to manipulate the physical properties of materials at nanoscale. By considering QShE in the design of nanostructures, it could be possible to suppress the unwanted effects and enhance the useful ones. Its size counterpart, QSE, have already been studied on many exotic systems such as topological insulators and superconductors \cite{Weis_2017,suprqsesc}. We may expect QShE to appear and make possibly important differences in many exotic systems from topological materials to superconductors as long as they are geometrically designed in an appropriate way.

QShE is a newly emerging field and open to further research. In addition to exploration of various exotic materials under QShE, fundamental investigations about its theory could be extended. Mode analysis in systems exhibiting QShE could also be worth to consider, as earlier studies on various typical systems reveal the importance of bound states in planar regions \cite{Trefethen2006} and quantum wavequides \cite{exner1,boundbook,exner2}. The research and models developed for the QShE may also have implications for the Casimir effect.

\bibliography{1Dref}
\bibliographystyle{unsrt}
\end{document}